\pdfoutput=1
\documentclass[10pt,journal,compsoc]{IEEEtran}

\usepackage{graphicx}
\usepackage{booktabs}
\usepackage{subcaption}
\usepackage{epstopdf}
\usepackage{diagbox}
\usepackage{threeparttable}
\usepackage{makecell, multirow,diagbox}

\usepackage[linesnumbered,ruled,vlined]{algorithm2e}
\usepackage{algpseudocode}
\usepackage{amssymb}
\usepackage{amsmath}
\usepackage{amsthm}
\usepackage{cite}
\usepackage{hyperref}
\usepackage{bm}
\usepackage{mathtools}
\usepackage{pgfmath}
\usepackage{dsfont}
\usepackage{xcolor}
\usepackage{color}
\usepackage{lineno}
\usepackage{colortbl}

\usepackage{flushend}

\def\ourmethod{LoRA-C}

\def\BibTeX{{\rm B\kern-.05em{\sc i\kern-.025em b}\kern-.08em
    T\kern-.1667em\lower.7ex\hbox{E}\kern-.125emX}}

\begin{document}
%
\title{\ourmethod{}: Parameter-Efficient Fine-Tuning of Robust CNN for IoT Devices}

\author{Chuntao~Ding,
        ~Xu~Cao,
        ~Jianhang~Xie,
        ~Linlin~Fan,
        ~Shangguang~Wang,
        ~Zhichao~Lu
\IEEEcompsocitemizethanks{
\IEEEcompsocthanksitem Chuntao Ding is with School of Artificial Intelligence, Beijing Normal University, Beijing, China.
E-mail: ctding@bnu.edu.cn

\IEEEcompsocthanksitem 
Xu Cao and Jianhang Xie are with Key Laboratory of Big Data \& Artificial Intelligence in Transportation, Ministry of Education, with School of Computer Science and Technology, Beijing Jiaotong University, Beijing 100044, China.
E-mail: \{caoxuu; xiejianhang\}@bjtu.edu.cn.

\IEEEcompsocthanksitem Linlin Fan is with the School of Computer and Information Engineering, Henan Normal University, Henan 453007, China.
E-mail: huazheng1106@gmail.com.

\IEEEcompsocthanksitem Shangguang Wang is with the State Key Laboratory of Networking and Switching Technology, Beijing University of Posts and Telecommunications, Beijing, China.
\protect E-mail: sgwang@bupt.edu.cn.

\IEEEcompsocthanksitem Zhichao Lu is with the Department of Computer Science, City University of Hong Kong, HKSAR.
E-mail: luzhichaocn@gmail.com.\\
}
}

\markboth{}%
{Shell \MakeLowercase{\textit{et al.}}: Bare Demo of IEEEtran.cls for IEEE Transactions on Magnetics Journals}

\IEEEtitleabstractindextext{%
\begin{abstract}
Efficient fine-tuning of pre-trained convolutional neural network (CNN) models using local data is essential for providing high-quality services to users using ubiquitous and resource-limited Internet of Things (IoT) devices.
Low-Rank Adaptation (LoRA) fine-tuning has attracted widespread attention from industry and academia because it is simple, efficient, and does not incur any additional reasoning burden.
However, most of the existing advanced methods use LoRA to fine-tune Transformer, and there are few studies on using LoRA to fine-tune CNN.
The CNN model is widely deployed on IoT devices for application due to its advantages in comprehensive resource occupancy and performance.
Moreover, IoT devices are widely deployed outdoors and usually process data affected by the environment (such as fog, snow, rain, etc.).
The goal of this paper is to use LoRA technology to efficiently improve the robustness of the CNN model.
To this end, this paper first proposes a strong, robust CNN fine-tuning method for IoT devices, \ourmethod{}, which performs low-rank decomposition in convolutional layers rather than kernel units to reduce the number of fine-tuning parameters.
Then, this paper analyzes two different rank settings in detail and observes that the best performance is usually achieved when ${\alpha}/{r}$ is a constant in either standard data or corrupted data.
This discovery provides experience for the widespread application of \ourmethod{}.
Finally, this paper conducts many experiments based on pre-trained models.
Experimental results on CIFAR-10, CIFAR-100, CIFAR-10-C, and Icons50 datasets show that the proposed \ourmethod{}s outperforms standard ResNets.  
Specifically, on the CIFAR-10-C dataset, the accuracy of \ourmethod{}-ResNet-101 achieves 83.44\% accuracy, surpassing the standard ResNet-101 result by +9.5\%.
On the Icons-50 dataset, the accuracy of \ourmethod{}-ResNet-34 achieves 96.9\% accuracy, surpassing the standard ResNet-34 result by +8.48\%.
In addition, compared with full parameter fine-tuning, \ourmethod{} can reduce the amount of updated parameters by more than 99\%.
\end{abstract}
}

\maketitle

\IEEEpeerreviewmaketitle

\section{Introduction} \label{ref-introduction}
\IEEEPARstart{T}{raining} large models is time-consuming and expensive.
For example, training Meta's LLaMA-65B model requires 2,048 NVIDIA A100 GPUs, takes about 21 days, and costs more than \$ 2.4 million.
Larger models, such as OpenAI's GPT-3, contain 175 billion parameters and cost more than \$ 4 million to train\footnote{https://www.cnbc.com/2023/03/13/chatgpt-and-generative-ai-are-booming-but-at-a-very-expensive-price.html}.
Moreover, these large models~\cite{Touvron@LLaMA, Chowdhery@PaLM} usually have broad general capabilities, but their performance on specific tasks is insufficient to meet practical needs.
Using task-specific data to fine-tune these large models to meet the needs of specific tasks has attracted widespread attention and spawned many effective fine-tuning methods~\cite{Ding@Parameter, Houlsby@Parameter, Jonas@AdapterFusion, Andreas@AdapterDrop, Mahabadi@Compacter, Li@PrefixTuning, Hu@LoRA, Chen@AdaptFormer, Lian@Scaling, Fan@DPOK}.
Among these fine-tuning methods, Low-Rank Adaptation (LoRA)~\cite{Hu@LoRA, Kopiczko@VeRA,  Dettmers@QLoRA, Zhang@Adaptive, Li@LoftQ, Zhang@LoRAPrune, Liu@DoRA} is widely used in the Transformer architecture because of its simplicity, effectiveness, and no additional reasoning burden.
\begin{figure}
\centering
\includegraphics[width=1\linewidth]{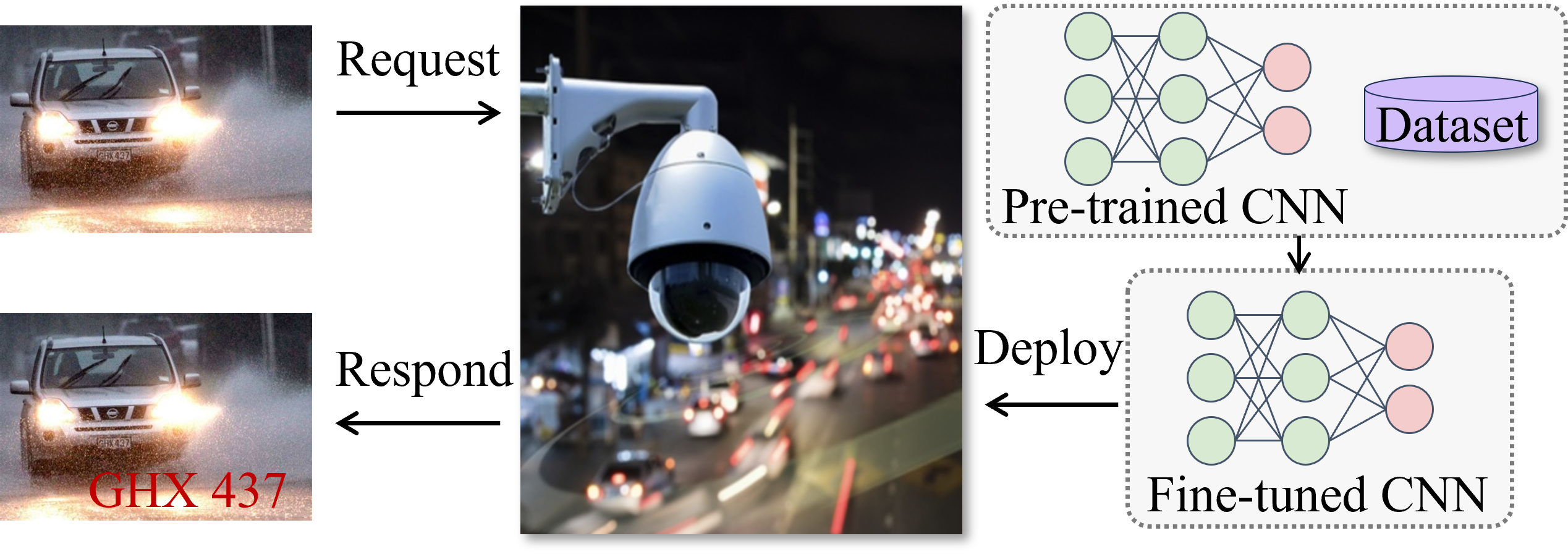}
\caption {Overview of system architecture.}
\label{fig:overview_problemState}
\end{figure} 

On the other hand, ABI Research reports that the global number of smart cameras using artificial intelligence (AI) chips will reach 350 million in 2025~\footnote{https://www.abiresearch.com/press/global-installed-base-smart-city-cameras-ai-chipset-reach-over-350-million-2025/}.
In addition, Internet of Things (IoT) devices are widely deployed outdoors and usually process data affected by the environment (such as fog, snow, rain, etc.).
Compared with Transformer~\cite{Vaswani@Attention, Liu@Swim, Liu@Swimv2}, Convolutional Neural Networks (CNN)~\cite{Karen@Very, Zhang@ShuffleNet, Huang@Densely, resnet} have more advantages in balancing resource consumption and performance in visual tasks and has been widely deployed on resource-constrained IoT devices.
Hence, it has become mainstream to use local data to fine-tune pre-trained CNN models transmitted from the cloud to achieve high performance on ubiquitous smart cameras containing AI chips.
As shown in Fig.~\ref{fig:overview_problemState}, the pre-trained model is fine-tuned using local data on IoT devices to improve the model's performance in processing images affected by the environment and provide high-quality services.

However, there are few studies on fine-tuning CNN using LoRA.
The goal of this paper is to use LoRA to fine-tune the CNN model on IoT devices to improve its robustness and provide high-quality services.

To this end, this paper first proposes the \ourmethod{}, a layer-wise LoRA for CNN fine-tuning.
Inspired by the application of LoRA in the Transformer structure, a natural approach is to use the low-rank decomposition of each convolution kernel as the newly added weights.
However, due to the kernel parameters sharing characteristics of CNN, low-rank decomposition based on convolution kernels will result in too many updated parameters, making it difficult to deploy on resource-constrained IoT devices.
Therefore, instead of adding low-rank decomposition kernel-wise, this paper proposes \ourmethod{}, which adds low-rank decomposition in convolutional layer-wise to efficiently fine-tune CNN models.
This paper also experimentally verifies that \ourmethod{} achieves significant performance improvements with only a few parameters updated.

Then, this paper discusses in detail the relationship between $\alpha$, which controls the weight of the newly added parameters, and the rank of $r$ of the matrix.
Extensive experiments show that the proposed \ourmethod{} usually performs best when $\alpha/ r$ is a constant. This finding provides experience for efficiently fine-tuning CNN models on IoT devices.
This paper also studies two settings of matrix rank $r$ and gives rank $r$ settings suitable for different models through many experiments.

Finally, this paper conducts extensive experiments on the CIFAR-10, CIFAR-100, CIAFR-10-C, and Icons-50 datasets, and the experimental results show that our proposed \ourmethod{} achieves strong performance. 
For example, our proposed \ourmethod{}-ResNet-50 achieves 82.98\% accuracy on the CIFAR-100 dataset, which surpasses the standard ResNet-50 result by +5.29\%.
On the CIFAR-10-C and Icons-50 datasets, our proposed \ourmethod{}-ResNet-34 achieves an average accuracy of 78.45\% and 96.9\%, respectively, which is +5.9\% and +8.48\% higher than the results of the standard ResNet-34.
In addition, we also show that \ourmethod{} has the advantage in the number of updated parameters, making it more suitable for deployment on resource-constrained IoT devices. 

In summary, our main contributions are as follows:
\begin{itemize}
    \item To our knowledge, this is the first work that seeks to improve CNN robustness by incorporating LoRA. Specifically, according to the characteristics of the CNN model, this paper proposes \ourmethod{}, a convolutional layer-wise low-rank decomposition method for effective fine-tuning.
    
    \item This paper observes that the proposed \ourmethod{} generally performs best when $\alpha/ r$ is a constant, which provides experience for efficiently fine-tuning CNN models on IoT devices.

    \item This paper compares the impact of rank settings on model robustness and analyzes its relationship with model capacity in detail.
    \item Extensive experiments on the CIFAR-10, CIFAR-100, CIFAR-10-C, and Icons-50 datasets demonstrate that our proposed \ourmethod{} achieves higher accuracy.
\end{itemize}

The remainder of the paper is organized as follows. Section~\ref{ref:2-related} reviews the work on convolutional neural networks, cloud-device collaboration, and parameter-efficient fine-tuning technologies. Section~\ref{ref-proposedapproach} describes the proposed \ourmethod{} in detail. Section~\ref{ref-experiments} presents our evaluation results, and Section~\ref{ref-conclusion} concludes the paper.

\section{Related Work}\label{ref:2-related} 
This section briefly introduces the Convolutional Neural Networks (CNN), cloud-device collaboration, and Parameter-Efficient Fine-Tuning (PEFT) techniques.
\subsection{Convolutional Neural Networks}
The great success of convolution operations in visual tasks has spawned many classic convolutional neural network (CNN) models in recent years, such as AlexNet~\cite{Alex@ImageNet}, VGG~\cite{Karen@Very}, Inception and its variants~\cite{Szegedy@Going, Ioffe@Batch, Szegedy@Rethinking, Szegedy@Inception}, ResNet and its variants~\cite{resnet, He@Identity}, Xception~\cite{Chollet@Xception}, DenseNet~\cite{Huang@Densely}, SENet~\cite{Hu@Squeeze}, etc.
Among them, ResNet~\cite{resnet} turns the CNN model into a deep CNN, introducing the residual module.
SENet~\cite{Hu@Squeeze} further improves the performance of the CNN model by introducing the idea of attention in the channel dimension.

The lightweight CNN models include the MobileNets series~\cite{Howard@MobileNets, mobilenetv2, Howard@Searching}, which separates spatial and channel dimensions; the ShuffleNet series~\cite{Zhang@ShuffleNet, Ma@ShuffleNet}, which uses group convolution; and the GhostNet series~\cite{han2020ghostnet, Tang@GhostNetV2}, which uses linear transformation to generate other feature maps based on a few feature maps.
Recently, RepVGG~\cite{Ding_2021_CVPR} introduces the re-parameterization to speed up the inference of the CNN model.
MonoCNN~\cite{Ding@Towards,Yu@Localized} and SineFM~\cite{Lu@Seed} have been proposed to improve the robustness of CNN models.
On the one hand, they reduce the number of learning parameters in the CNN model to alleviate the parameter update from being completely dependent on training data. On the other hand, they use nonlinear transformation rules to regularize the model to improve the robustness of the CNN model.

Different from the above methods, this study provides another idea, which aims to explore the use of LoRA to improve the robustness of the CNN model, so as to help CNN provide high-quality services on IoT devices.

\begin {figure*}[ht]
\centering
\includegraphics[width=1\linewidth]{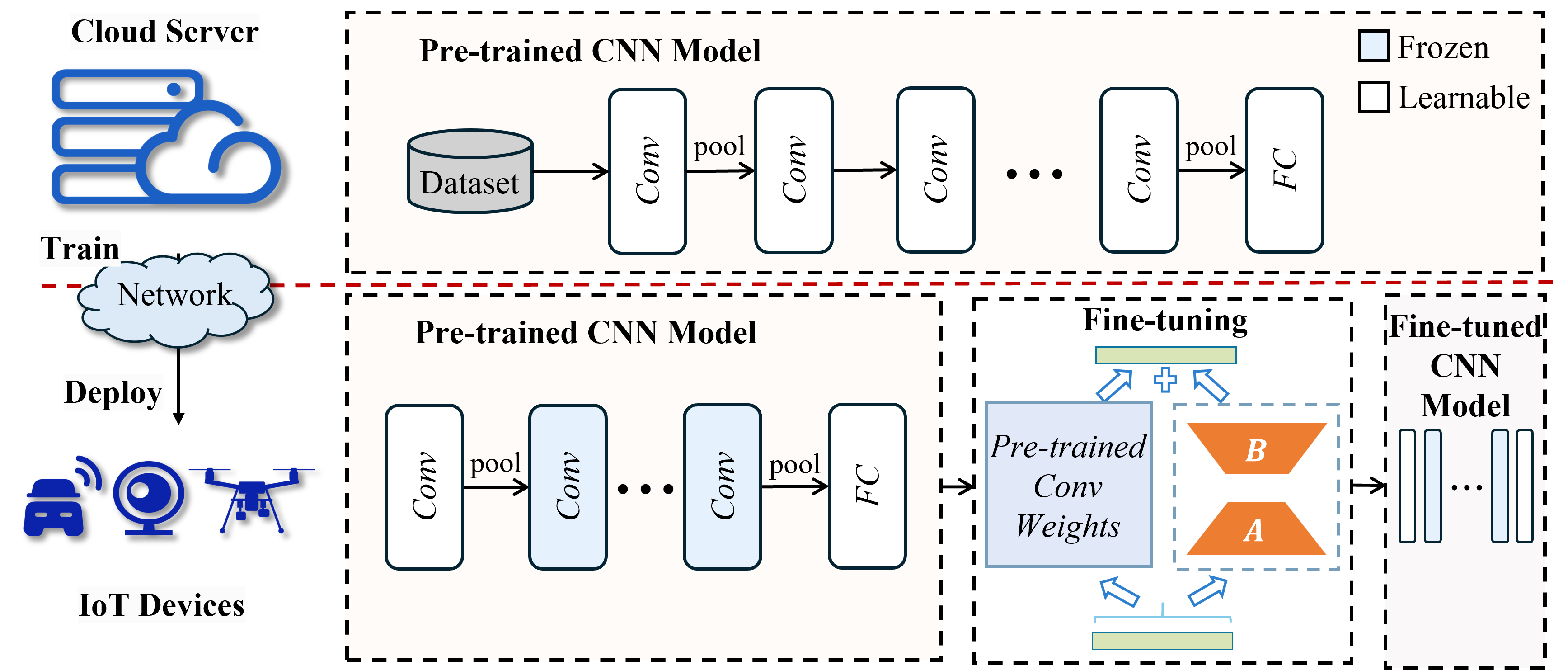}
\caption {Overview of the proposed framework.} 
\label{fig:lorac_framework}
\end{figure*} 
\subsection{Cloud-Device Collaboration}
Due to the limited resources of IoT devices, cloud-device collaborative training and deployment of CNN models have become mainstream.
For example, Kang \emph{et al.}~\cite{Kang@Neurosurgeon} propose to divide the CNN into a head running on the device and a tail running on the cloud and determine the split point based on the device load and the amount of cloud-to-device parameter transmission.
Zhang~\emph{et al.}~\cite{Zhang@Collaborative} train the CNN model through cloud-edge collaboration and prune the CNN in the cloud to minimize the number of model transmission parameters while retaining the model performance. 
Stefanos~\emph{et al.}~\cite{Stefanos@SPINN} proposed a progressive inference method for collaborative device and cloud computing and used compression~\cite{Han@Deep} to reduce the amount of parameter exchange between the device and the cloud. 
However, the above methods are highly dependent on network conditions.
When the network condition is unstable or unavailable, the service quality will be reduced or even unavailable.

To decouple service quality from network conditions, many works proposed that the cloud server first trains the model and then sends the trained model to IoT devices~\cite{Ding@Towards, Yu@Localized, Lu@TFormer, Lu@Seed, Ding@A}.
For example, Ding \emph{et al.}~\cite{Ding@Towards} propose to train a CNN model containing a small number of learning parameters, namely MonoCNN, on the cloud, and then send the trained MonoCNN model to IoT devices.

This paper studies the cloud training of CNN models first and then sends the trained CNN models to IoT devices. 
The difference is that IoT devices efficiently fine-tune the CNN models sent from the cloud based on their data rather than directly using these CNN models to provide services.

\subsection{Parameter-Efficient Fine-Tuning Techniques}
In recent years, many excellent parameter-efficient fine-tuning techniques~\cite{Mao@UniPELT, Houlsby@Parameter, Ding@Parameter} have been proposed, aiming to improve the performance of pre-trained models on new tasks by minimizing the number of fine-tuning parameters and computational complexity, thereby reducing the cost of training models on new tasks. Existing methods can be divided into three categories.

The first category is called the Adapter~\cite{Houlsby@Parameter, Jonas@AdapterFusion, Andreas@AdapterDrop, Mahabadi@Compacter}, which inserts a module with a small number of parameters into the pre-trained model for each downstream task, and the inserted module with a small number of parameters is called the Adapter module.
In this category, only the Adapter module parameters are updated during training for downstream tasks, thereby efficiently migrating the capabilities of the powerful foundation models to many downstream tasks while ensuring the model's performance in downstream tasks. 
However, it increases the depth of the pre-trained model, thereby increasing the latency of model reasoning.

The second category is called the P-tuning~\cite{Li@PrefixTuning, Lester@The, Liu@P-Tuningv2}, which constructs a task-related virtual token as a prefix before inputting the token. 
It then only updates the parameters of the prefix part during training, while the other parameters in the Transformer are fixed.
However, P-tuning methods are challenging to train and reduce the available sequence length of the model.

The third category is called LoRA and its variants~\cite{Hu@LoRA, Dettmers@QLoRA, Zhang@Adaptive, Li@LoftQ, Zhang@LoRAPrune, Liu@DoRA}. The full name of LoRA is low-rank adaptation, and is first proposed by Hu~\emph{et al.}~\cite{Hu@LoRA}.
Its core idea is to simulate the change of model parameters when the model adapts to new tasks through low-rank decomposition to realize indirect training of large models with an extremely small number of parameters.
LoRA assumes that the pre-trained foundation models are over-parameterized and have a small ``intrinsic dimension''~\cite{Li@Measuring, Aghajanyan@Intrinsic}, that is, there is an extremely low dimensional.
Fine-tuning the parameters of the extremely low dimension can have the same effect as fine-tuning in the full parameter space.
LoRA does not change the depth of the pre-trained model, and can merge parameters added for new tasks during inference, thus generating no additional inference delay.
Given the above advantages of LoRA, LoRA and its variants have been extensively studied in many applications. 
For example, Zhang \emph{et al}.~\cite{Zhang@LoRAPrune} combine LoRA with pruning and propose LoRAPrune.
Dettmers \emph{et al.}~\cite{Dettmers@QLoRA} and Li \emph{et al.}~\cite{Li@LoftQ} combine LoRA with quantization and propose QLoRA and LoftQ, respectively.
Liu \emph{et al.}~\cite{Liu@DoRA} propose the DoRA based on LoRA. It first decomposes the pre-trained weights into their amplitude and direction components and then fine-tunes the two, which enhances the learning ability and training stability of LoRA.

Besides, the research on LoRA fine-tuning with CNN model is rare currently~\cite{Aleem@ConvLoRA, aggarwal2024advancing, zhong2024convolution, yeh2024navigating}.
The former two works~\cite{Aleem@ConvLoRA, aggarwal2024advancing} attempted to combine LoRA with convolution but did not analyze it further and deeper.
Zhong \emph{et al.}~\cite{zhong2024convolution} combined the lightweight convolution into the mixture of expert network to improve LoRA, but did not fine-tune the convolution itself.
Yeh \emph{et al.}~\cite{yeh2024navigating} proposed the LoCon which has two convolutions for dimension-down (with $k\times k$ convolution) and dimension-up (with $1\times1$ convolution) to simulate the matrices decomposition to reduce the number of fine-tuning parameters.

Our research falls within the third category. 
We aim to use LoRA to improve the robustness of CNN models, a direction with high practical application value but little research.
This paper will study in depth how to apply LoRA to CNN, so that it can effectively improve the performance of the CNN model while only updating a very small number of parameters.

\section{Design of the Proposed Approach} \label{ref-proposedapproach}

\subsection{Overview}

Fig.~\ref{fig:lorac_framework} illustrates the overview of the proposed approach.
IoT devices are usually deployed outdoors and collect a lot of data affected by the environment (such as fog, rain, and snow).
We refer to the performance of the CNN model when processing this type of data as the robustness of the model.
The standard CNN model is low-robust when processing this type of data.
We aim to efficiently fine-tune the pre-trained model on resource-constrained IoT devices using local data to improve model robustness and handle the environmental-affected data collected by IoT devices.

To this end, we propose parameter-efficient fine-tuning of robust CNN method, \ourmethod{}.
We first use the cloud to train the complex CNN model, known as the pre-trained model, and then send the pre-trained model to IoT devices.
On the IoT device, we freeze all convolutions of the pre-trained model, excluding its first and last layers, and add \ourmethod{} branches to each frozen convolution layer.
Then, the IoT device uses a small amount of local data to fine-tune this pre-trained model based on the \ourmethod{}.
When providing CNN-powered IoT smart services, the parameters of the \ourmethod{} branch can be fused into the pre-trained convolutional weights, which will not introduce additional inference latency.

Next, we will conduct a fine-grained analysis from motivation and design of \ourmethod{}.

\subsection{Motivation}
This section will illustrate our motivation by analyzing the two most relevant techniques, namely low-rank adaptation and the convolutional layer.

\noindent{\bf{Low-Rank Adaptation}}~\cite{Hu@LoRA}:
Low-rank adaptation (LoRA) uses low-rank decomposition to simulate the change in model parameters when the model adapts to new tasks, thereby achieving indirect training of large models by updating very few parameters. Formally, for a pre-trained weight matrix $\bm{W}_0\in \mathbb{R}^{d\times k}$, the newly added weight matrix $\Delta \bm{W} = \bm{B}\bm{A} \in \mathbb{R}^{d\times k}$, matrix $\bm{A}\in \mathbb{R}^{r\times k}$ and matrix $\bm{B}\in \mathbb{R}^{d\times r}$, and the rank $r\ll min(d,k)$.
When fine-tuning the model, only the parameters of $\Delta \bm{W}$ are updated, while the parameters of $\bm{W}_0$ are frozen.
$\bm{W}_0$ and $\Delta \bm{W}$ are multiplied with the same input, and their respective output vectors are summed coordinate-wise.
Given input $x$, the forward pass yields:
\begin{equation}
\begin{array}{l}
h = \bm{W}_0x + \alpha \Delta \bm{W}x
\end{array},
\label{eq:9}
\end{equation}
where $\alpha$ is a constant. In the $\Delta \bm{W}$, the settings of LoRA initialize $\bm{A}$ with the random Gaussian distribution and set $\bm{B}$ to all zeros.
Since $r\ll min(d,k)$, the number of parameters of $\Delta \bm{W}$ is much smaller than that of $\bm{W}_0$.
In addition, during inference, $\Delta \bm{W}$ can be completely added to $\bm{W}_0$ without changing the model architecture and without incurring any additional inference overhead.

\noindent{\bf{Convolutional Layer}}~\cite{LeCun@Backpropagation, LeCun@Gradient, Alex@ImageNet}:
The convolution uses sliding kernels to extract features.
Let $x\in \mathbb{R}^{c_\mathrm{in}\times w\times h}$ denote the input feature maps, where $w$, $h$, and $c_\mathrm{in}$ are the input width, input height, and input channels, respectively; 
and $\bm{W}\in \mathbb{R}^{c_\mathrm{out}\times c_\mathrm{in}\times k\times k}$ denote the convolution weights, where a kernel with size of $k\times k$.
Then the generated output feature maps is $y\in \mathbb{R}^{c_\mathrm{out}\times w'\times h'}$, where $w'$, $h'$ and $c_\mathrm{out}$ are the output width, output height, and output channels, respectively.
So the forward of the convolutional layer can be described by:
\begin{equation}
\begin{array}{l}
y = \bm{W}\otimes x
\end{array},
\label{eq:conv}
\end{equation}
where the $\otimes$ is a convolution operation.
For a kernel $\bm{W}^{(n)}\in\mathbb{R}^{c_\mathrm{in}\times k\times k}, (0 < n < c_\mathrm{out})$, the output feature map $y^{(n)}$ in output channel $n$ is:
\begin{equation}
\begin{array}{l}
y^{(n)} = {\bm{W}^{(n)}\otimes x}
\end{array},
\label{eq:kernel}
\end{equation}
where a value of feature map $y^{(n)}$ in the position $(i,j)$ can be calculated by:
\begin{equation}
\begin{array}{l}
y_{i,j}^{(n)} = \sum\limits_{m=1}^{c_\mathrm{in}}\sum\limits_{u=1}^{k}\sum\limits_{v=1}^{k}{\bm{W}^{(n)}_{m,u,v}\cdot x_{m,i-u+1,j-v+1}}
\end{array}.
\label{eq:kernel_value}
\end{equation}

Most existing studies use LoRA to fine-tune models based on the Transformer architecture, and there are very few LoRA studies targeting CNN.
Given the wide application of CNN on IoT devices, this paper studies using LoRA to fine-tune the CNN model.
When fine-tuning the Transformer-based architecture, the $\bm{B}\bm{A}$ in LoRA acts as the $\bm{Q}$, $\bm{K}$, and $\bm{V}$ matrices.
{\emph{So how can LoRA be efficiently combined with CNN?}}
\begin{figure}
\centering
\includegraphics[width=1\linewidth]{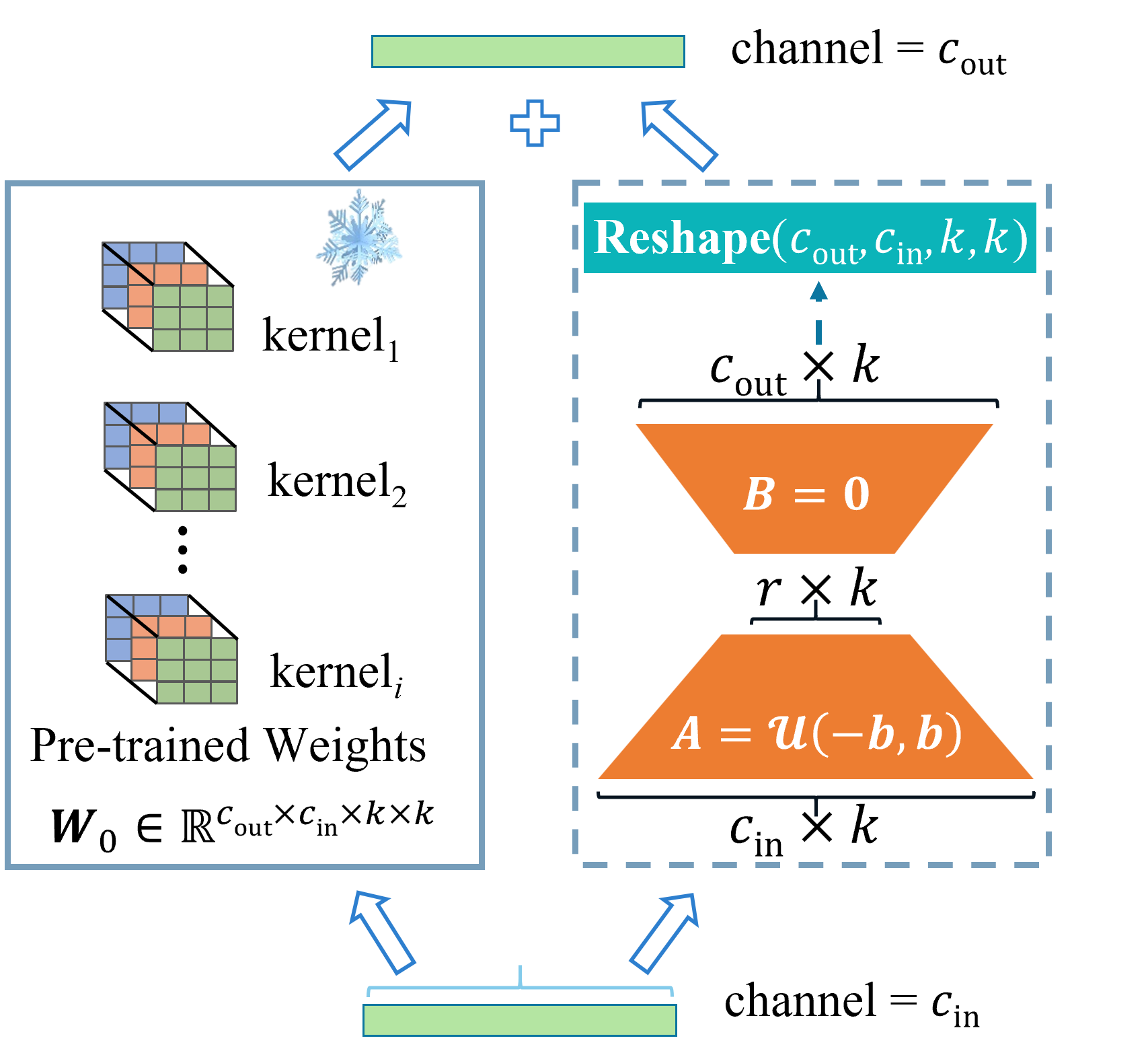}
\caption {The proposed \ourmethod{}.}
\label{fig:loracnn}
\end{figure} 

\subsection{Design of \ourmethod{}}
This section first presents \ourmethod{} and then explains the reasons for this design.

\noindent{\bf{\ourmethod{}}}:
Fig.~\ref{fig:loracnn} illustrates the proposed \ourmethod{}, which is a convolutional layer-wise low-rank decomposition method for CNN fine-tuning.
For a convolutional layer with the input channel $c_\mathrm{in}$ and output channel $c_\mathrm{out}$, the pre-trained convolutional weights $\bm{W}_0 \in \mathbb{R}^{c_\mathrm{out}\times c_\mathrm{in}\times k\times k}$ are froze by removing the gradients, which are not updated during fine-tuning.

We introduce a LoRA-C branch with layer-wise decomposed matrices for updating the pre-trained weights $\bm{W}_0$, that is, decomposing the updating increment of weights $\Delta \bm{W}$ into two matrices, $\bm{A}$ and $\bm{B}$, where $\bm{A}\in \mathbb{R}^{r\times c_\mathrm{in}\times k}$ and $\bm{B}\in \mathbb{R}^{c_\mathrm{out}\times k\times r}$.

As with LoRA, we also introduce a constant, $\alpha$, to scale the weight of the newly added parameters in LoRA-C, and the updating of fine-tuning can be described by Equation~\ref{eq:lora_update}:
\begin{equation}
\begin{array}{l}
\bm{W}=\bm{W}_0+\alpha\Delta \bm{W}=\bm{W}_0+\alpha\bm{B}\bm{A}
\end{array}.
\label{eq:lora_update}
\end{equation}

The increments of pre-trained weights $\bm{W}_0$ brought by fine-tuning is equivalent to $\bm{B}\bm{A}$, where $\bm{B}\bm{A}\in \mathbb{R}^{c_\mathrm{out}\times k\times c_\mathrm{in}\times k}$.
The $\bm{B}\bm{A}$ matrix can be reshape into a tensor whose dimensions are equal to the pre-trained weights $\bm{W}_0$, i.e., $\bm{\mathrm{Reshape}}\left(\bm{B}\bm{A}\right) \in \mathbb{R}^{c_\mathrm{out}\times c_\mathrm{in}\times k\times k}$.

For the initialization of $\bm{A}$ and $\bm{B}$ matrices, we follow the settings of LoRA, set the matrix $\bm{B}$ with all zeros $\mathcal{Z}$ and initialize matrix $\bm{A}$ with Kaiming uniform distribution $\mathcal{U}(-b, b)$, where the $b$ is the bounding of uniform distribution~\cite{he2015delving}. 
Let $\bm{T}$ denote a random initial tensor, the initialization of $\bm{A}$ and $\bm{B}$ is as follows:
\begin{equation}
\begin{array}{l}
\left\{
\begin{aligned}
\bm{A} &= \mathcal{U}(\bm{T}),\ \bm{T}\in \mathbb{R}^{r\times c_\mathrm{in} \times k}\\
\bm{B} &= \mathcal{Z}(\bm{T}), \ \bm{T}\in \mathbb{R}^{c_\mathrm{out} \times k\times r}
\end{aligned}
\right.
\end{array}.
\label{eq:init_ab_layer_wise}
\end{equation}
The forward of convolution operation in LoRA-C is: 
\begin{equation}
\begin{array}{l}
y=\left(\bm{W}_0+\alpha \bm{B}\bm{A}\right)\otimes x
\end{array}.
\label{eq:forward_layer_wise}
\end{equation}

For the rank of $\Delta \bm{W}$ (or $\bm{B}\bm{A}$), we provide two settings of $r$ and $r\ast k$, while the impact of the two different ranks will be analyzed in Section~\ref{ref-rank-analysis}.

\subsubsection{Decomposed Matrices Granularity}
To apply the mechanism of LoRA fine-tuning to CNN model, we made modifications in layer-wise granularity to adapt to the structure of convolutional layer. 
\emph{So why choose a laye-wise decomposition matrices instead of the more fine-grained kernel-wise?}

We will analyze the choice of layer-wise decomposition matrices granularity in terms of parametric efficiency.
For the full fine-tuning, the number of updated parameters for pre-trianed weights $\bm{W}_0$ is:
\begin{equation}
\begin{array}{l}
\mathcal{P}_\mathrm{full\_ft}=c_\mathrm{out}\cdot c_\mathrm{in}\cdot k^2
\end{array}.
\label{eq:update_fft}
\end{equation}

We aim to reduce the number of updated parameters in the convolutional weights by utilizing the LoRA technique.

\noindent{\emph{Kernel-Wise Decomposed Matrices}}: Based on the previous experience of LoRA in attention, we first consider assigning the $\bm{B}\bm{A}$ matrix for each convolutional kernel to fine-tune, i.e., kernel-wise decomposed matrices in LoRA-C.

Let $\bm{W}^{\left(i\right)}\in
\mathbb{R}^{c_\mathrm{in}\times k\times k}$ denote the $i$-th convolutional kernel in output channel $i$, where $0 < i< c_\mathrm{out}$.
So, for each $\bm{W}^{\left(i\right)}$, it can introduce a pair of low-rank matrices $\bm{A}^{\left(i\right)}$ and $\bm{B}^{\left(i\right)}$ with rank $r$, where $\bm{A}^{\left(i\right)}\in \mathbb{R}^{r\times c_\mathrm{in}\times k}$ and $\bm{B}^{\left(i\right)}\in \mathbb{R}^{k\times r}$. 
So the updating of one kernel can be given by Equation~\ref{eq:kernel_wise_lorac}:
\begin{equation}
\begin{array}{l}
\bm{W}^{\left(i\right)}=\bm{W}_0^{\left(i\right)}+\alpha\Delta \bm{W}^{\left(i\right)}=\bm{W}_0^{\left(i\right)}+\alpha\bm{B}^{\left(i\right)}\bm{A}^{\left(i\right)}
\end{array},
\label{eq:kernel_wise_lorac}
\end{equation}
so the forward of convolution in kernel-wise is:
\begin{equation}
\begin{array}{l}
y^{\left(i\right)}=\left(\bm{W}_0^{\left(i\right)}+\alpha \bm{B}^{\left(i\right)}\bm{A}^{\left(i\right)}\right) \otimes x
\end{array}.
\label{eq:forward}
\end{equation}

For kernel-wise decomposition, the number of parameters for $\bm{A}^{\left(i\right)}$ and $\bm{B}^{\left(i\right)}$ is $c_\mathrm{in}\cdot r \cdot k+r \cdot k$, and there are $c_{out}$ kernels in a convolution layer. 
Thus, the number of updated parameters for kernel-wise decomposed matrices is:
\begin{equation}
\begin{array}{l}
\mathcal{P}_\mathrm{kernel\_wise}=c_\mathrm{out}\cdot (c_\mathrm{in}\cdot r\cdot k + r\cdot k)
\end{array}.
\label{eq:update_kernel_wise}
\end{equation}

\noindent{\emph{Layer-Wise Decomposed Matrices}}: 
For layer-wise decomposition using a pair of decomposed matrices with rank $r$ to fine-tune the convolutional weights in convolutional layer, the corresponding number of updated parameters can be described in Equation~\ref{eq:update_layer_wise}:
\begin{equation}
\begin{array}{l}
\mathcal{P}_\mathrm{layer\_wise}=(c_\mathrm{out}+c_\mathrm{in}) \cdot r \cdot k
\end{array}.
\label{eq:update_layer_wise}
\end{equation}

\noindent{\emph{Updated Parameters Reduction}}: For the two different LoRA-C granularities of low-rank decomposition, kernel-wise and layer-wise, we analyze the number of updated parameters compared to full fine-tuning, respectively. 

For an intuitive comparison, we choose the ratio of the amounts of updating parameter of two LoRA-C granularities to the full fine-tuning number of parameters.
The ratio of kernel-wise decomposition can be calculated via Equation~\ref{eq:ratio_kw}.
\begin{equation}
\begin{array}{l}
\begin{aligned}
\mathcal{R}_\mathrm{kernel\_wise} &= \frac{\mathcal{P}_\mathrm{kernel\_wise}}{\mathcal{P}_\mathrm{full\_ft}} = \frac{c_\mathrm{out} \cdot (c_\mathrm{in} \cdot r \cdot k + r \cdot k)}{c_\mathrm{out} \cdot c_\mathrm{in} \cdot k^2} \\
 &= \frac{c_\mathrm{in} \cdot r+r}{c_\mathrm{in} \cdot k}
\end{aligned}
\end{array}.
\label{eq:ratio_kw}
\end{equation}
We can find that the number of parameters in kernel-wise will be greater than the full fine-tuning one when $r \geq k$ as $\mathcal{R}_\mathrm{kernel\_wise} \geq \frac{c_\mathrm{in} \cdot k+k}{c_\mathrm{in} \cdot k} > 1$. 
For an example, the representative CNN, ResNet, the kernel size $k$ typically takes values of 1 or 3, so the $r$ should not be greater than 3 for a positive gain.
However, the model usually has ill perform when $r$ is too small as our ablation studies in Section~\ref{ref-hyperparameter}.
Therefore, the design of kernel-wise LoRA-C is difficult to achieve a balance between performance and parametric efficiency.

The ratio of layer-wise decomposition is:
\begin{equation}
\begin{array}{l}
\begin{aligned}
\mathcal{R}_\mathrm{layer\_wise} &= \frac{\mathcal{P}_\mathrm{layer\_wise}}{\mathcal{P}_\mathrm{full\_ft}} = \frac{(c_\mathrm{out}+c_\mathrm{in}) \cdot r \cdot k}{c_\mathrm{out} \cdot c_\mathrm{in} \cdot k^2} \\
 &= \frac{(c_\mathrm{out}+c_\mathrm{in}) \cdot r}{c_\mathrm{out} \cdot c_\mathrm{in} \cdot k}
\end{aligned}
\end{array}.
\label{eq:ratio_lw}
\end{equation}
In the layer-wise granularity, the number of parameters mainly depends on $c_\mathrm{out},c_\mathrm{in},r$ and $k$.
In typical CNN for computer vision, $c_\mathrm{out}$ and $c_\mathrm{in}$ are generally greater than 3 (or even larger, e.g., 64, 128, 256, ...), so $c_\mathrm{out}+c_\mathrm{in}$ is usually smaller than $c_\mathrm{out} \cdot c_\mathrm{in}$.
For example, for a $k = 1$ convolution with $c_\mathrm{in}=256$, $c_\mathrm{out}=512$, and rank $r = 128$, the $ \mathcal{R}_\mathrm{layer\_wise} = \frac{(512+256) \cdot 128}{512 \cdot 256 \cdot 1} = 0.75 < 1$ can also maintain a positive gain of updated parameters reduction in a higher rank $r$ ($r=128 \gg k=1$).

\begin{figure*}[t]
    \centering
    \begin{subfigure}[b]{0.245\textwidth}
    \centering
    \includegraphics[width=1\textwidth]{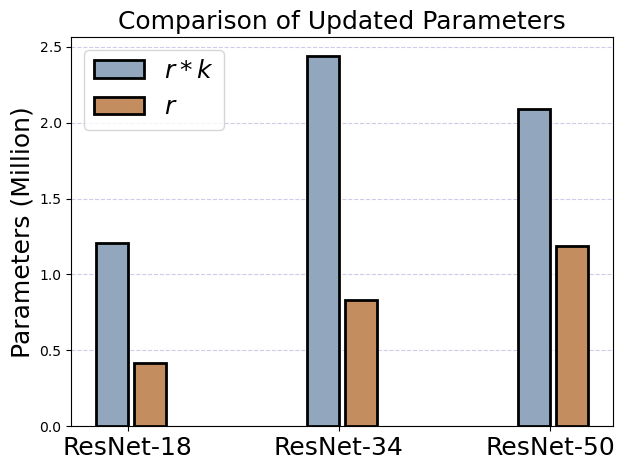}
    \caption{\label{subfig:parameters}}
    \end{subfigure}\hfill
    \centering
    \begin{subfigure}[b]{0.245\textwidth}
    \centering
    \includegraphics[width=1\textwidth]{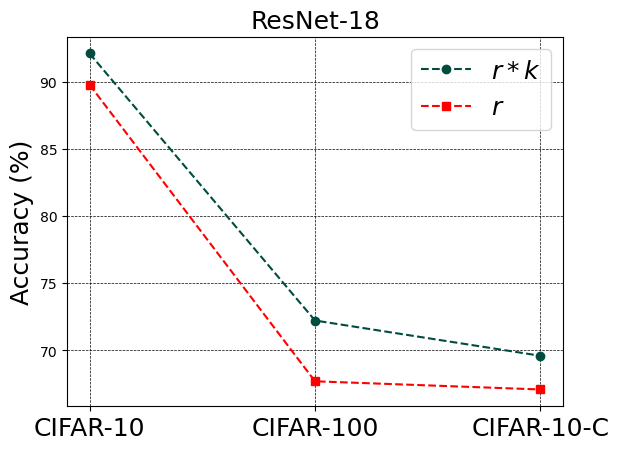}
    \caption{\label{subfig:resnet18_k_rk}}
    \end{subfigure}\hfill
    \centering
    \begin{subfigure}[b]{0.245\textwidth}
    \centering
    \includegraphics[width=1\textwidth]{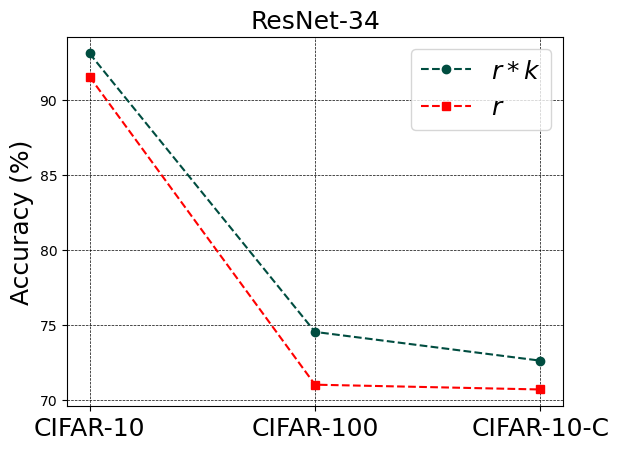}
    \caption{\label{subfig:resnet34_k_rk}}
    \end{subfigure}\hfill
    \centering
    \begin{subfigure}[b]{0.245\textwidth}
    \centering
    \includegraphics[width=1\textwidth]{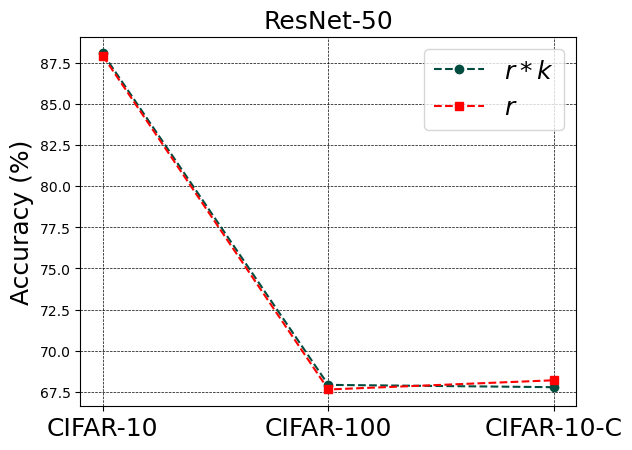}
    \caption{\label{subfig:resnet50_k_rk}}
    \end{subfigure}
    \caption{Comparison of updated parameters, performance, and robustness. (a) Comparison of updated parameters under two settings. (b) Performance comparison of the two settings when using ResNet-18. (c) Comparison of two settings when using ResNet-34. (d) Performance comparison of the two settings when using ResNet-50. We conclude that the second setting is used on models with model parameters greater than or equal to ResNet-50, and the first setting is used on models with model parameters less than ResNet-50.\label{fig:paravspervsrobust}}
\end{figure*}

Compared to kernel-wise decomposition which causes parameter inflation, the layer-wise one has fewer fine-tuning parameters for updating.
Therefore, we choose layer-wise decomposed matrices as the decomposition in LoRA-C.

\subsubsection{Impact of Decomposed Matrices Rank} \label{ref-rank-analysis}
There are also two settings of the rank of $\Delta \bm{W}$ when fine-tuning the CNN model using LoRA.
The first is to associate the rank with the size of the convolutional kernel, that is, the rank is $r\ast k$.
The second is to directly set rank to $r$.
The number of updated parameters for the latter is $1/k$ of the former.
{\emph{How to choose between these two settings?}}

To this end, on CIFAR-10/-100 datasets, based on ResNet-18/-34/-50, we set $r$ to $\{1, 2, 4, 8, 16, 32, 64\}$ and $\alpha$ to $\{1, 2, 4, 8, 16, 32, 64, 128\}$ for traversal.
The average amount of updated parameters for each model under different settings, the average performance of each model under each setting on each dataset, and the robustness on CIFAR-10-C dataset are shown in Fig.~\ref{fig:paravspervsrobust}.

As shown in Fig.~\ref{subfig:parameters}, the amount of updated parameters under the first setting is more than twice that of the second.
Correspondingly, as shown in Fig.~\ref{subfig:resnet18_k_rk} and Fig.~\ref{subfig:resnet34_k_rk}, the model accuracy under the first setting is higher than that under the second setting.
However, in the case of the ResNet-50, as shown in Fig.~\ref{subfig:resnet50_k_rk}, the model accuracy under the two settings is essentially equal, and even on CIFAR-10-C, the model under the second setting is more robust.

The two settings seem to be impossible to choose.
However, we see from Fig.~\ref{fig:paravspervsrobust} that the gap in accuracy between the two settings is narrowing when we go from ResNet-18 to ResNet-34 and then to ResNet-50. 
For example, on the CIFAR-10 dataset, when using ResNet-18, the model accuracy of the first setting is 2.33\% higher than that of the second one.
When using ResNet-34, the model accuracy of the first setting is 1.55\% higher than that of the second.
When using ResNet-50, the accuracy of the two settings is roughly the same.

It is worth mentioning that the second setting outperforms the first one in terms of improving model robustness as the model becomes larger in CIFAR-10-C.
For example, when using ResNet-18, the model accuracy of the first setting is 2.51\% higher than the model accuracy of the second one.
When using ResNet-34, the model accuracy of the first setting is 1.92\% higher than the model accuracy of the second one.
While using ResNet-50, the model accuracy of the second setting exceeds that of the first.

To this end, we comprehensively consider the number of updated parameters and model accuracy and conclude that the second setting is used on models with model parameters greater than or equal to ResNet-50, and the first setting is used on models with model parameters less than ResNet-50.
The experiments in this paper follow this setting.

\subsection{Procedures of \ourmethod{} on IoT Devices}

\subsubsection{\ourmethod{} Fine-Tuning}
The CNN fine-tuning procedure of \ourmethod{} consist of
two stages: preliminary of low-rank decomposition and tuning the pre-trained CNN model, as depicted in Algorithm~\ref{alg:lorac_ft_procs}.
\begin{algorithm}
\caption{LoRA-C Fine-Tuning}
\label{alg:lorac_ft_procs}
\KwIn{The pre-trained CNN model ${\bf{\mathcal{M}}}$, local corrupted datasets $\bf{\mathcal{D}}_\mathrm{local}$}
\KwOut{The robust CNN model ${\bf{\mathcal{M}}}_\mathrm{robust}$}

\textbf{Preliminary of Low-Rank Decomposition}:

Local corrupted datasets $\bf{\mathcal{D}}_\mathrm{local}$ initialization\;

\For{layer $\ell$ in ${\bf{\mathcal{M}}}$}{
    Set the rank $r$ of decomposed matrix $\Delta \bm{W}_\ell$\;
    Register LoRA-C branch with $\bm{A}_\ell$, $\bm{B}_\ell$ matrices as learnable parameters in layer-wise\;
    Freeze pre-trained convolutional weights $\bm{W}_\ell$ by removing gradients\;
    }

\textbf{Tuning the pre-trained CNN model ${\bf{\mathcal{M}}}$}:

\For{$epoch$ in $1,2, \ldots, epoch_\mathrm{max}$}{
    \For{$batch$ in $\bf{\mathcal{D}}_\mathrm{local}$}{
        $batch$ is a mini-batch from $\bf{\mathcal{D}}_\mathrm{local}$\;
        Activate LoRA-C branches with decomposed matrices $\bm{A}$, $\bm{B}$\;

        Compute predictions by ${\bf{\mathcal{M}}}$.forward()\;
        
        Compute gradients of matrices $\bm{A}$, $\bm{B}$\;
        
        Update parameters in LoRA-C branches\;
    }
}
\textbf{Return} ${\bf{\mathcal{M}}}_\mathrm{robust}$;
\end{algorithm}

\begin{figure*}[h]
\centering
\includegraphics[width=1\linewidth]{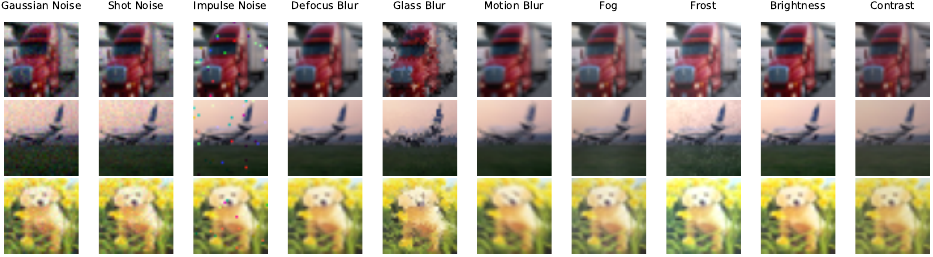}
\caption {Visualization examples of commonly observable corruptions in CIFAR-10-C dataset.}
\label{fig:CIFAR-10-C-data}
\end{figure*} 
\begin{figure}
\centering
\includegraphics[width=1\linewidth]{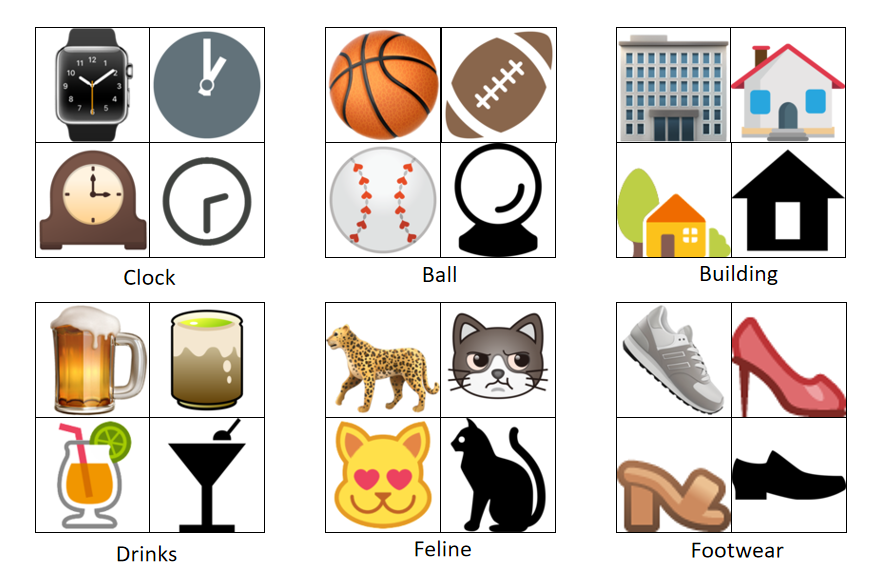}
\caption {Visualization examples of the Icons-50 dataset. For each category, the image in the upper left corner is from the test set, and the other three are from the training set. As shown, the style of training data is very different from that of test data.}
\label{fig:Icon-50-dataset}
\end{figure} 

\noindent{\emph{Preliminary of Low-Rank Decomposition}}: Before tuning, we have to prepare the settings of the pre-trained CNN model sent by the cloud server.
We need to add new branches to each convolution layer, referred to as the LoRA-C branch.
The LoRA-C branch initializes the $\bm{A}_\ell$ and $\bm{B}_\ell$ matrices at each layer $\ell$.
At the same time, the gradients need to be removed for each pre-trained convolutional weights and retained only for the LoRA-C branches to reduce the parameter burden of fine-tuning.

\noindent{\emph{Tuning the Pre-trained CNN Model}}: The on-device fine-tuning process is similar to model training, but usually using with fewer epochs.
The difference is that only the $\bm{A}$ and $\bm{B}$ matrices compute the gradients and update the parameters during tuning iteration.
Therefore, the small amount of computational overhead bringing by parametric efficiency enables the CNN fine-tuning of LoRA-C to adapt to the computational resources of IoT devices.

\subsubsection{\ourmethod{} CNN Inference}
After fine-tuning with local datasets, the IoT device can get a robust CNN model.
The pre-trained model can be removed if storage resources are insufficient, and only storing the updated LoRA-C branches.
This is because the pre-trained model can be sent by cloud assist and LoRA-C does not change pre-trained weights.
The IoT device only needs to maintain the portion of the incremental updates.
\begin{algorithm}
\caption{LoRA-C Inference}
\label{alg:lorac_infer_procs}
\KwIn{The robust CNN model ${\bf{\mathcal{M}}}_\mathrm{robust}$}
\KwOut{The robust CNN inference model ${\bf{\mathcal{M}}}_\mathrm{inference}$}

\textbf{Convert the ${\bf{\mathcal{M}}}_\mathrm{robust}$ to the ${\bf{\mathcal{M}}}_\mathrm{inference}$}:

\For{layer $\ell$ in ${\bf{\mathcal{M}}}_\mathrm{robust}$}{
    Recompose $\bm{A}_\ell$, $\bm{B}_\ell$ matrices to $\Delta \bm{W}_\ell$\;
    Merge the updated incremental matrix $\Delta \bm{W}_\ell$ into frozen pre-trained conv weights $\bm{W}_\ell$ by coordinate-wise summation\;
    Streamline this layer further with other re-parameterization techniques, e.g., BN fusion\;
    }
\textbf{Return} ${\bf{\mathcal{M}}}_\mathrm{inference}$;
\end{algorithm}

For on-device CNN model inference, as shown in Algorithm~\ref{alg:lorac_infer_procs}, we only need to recompose the decomposed matrices $\bm{A}_\ell$ and $\bm{B}_\ell$ of the LoRA-C branches into the increments $\Delta \bm{W}_\ell$ in each layer $\ell$, then fuse them to pre-trained weights by coordinate-wise summation to eliminate the branches.
Finally, we can streamline the model further by some other re-parameterization techniques~\cite{Ding_2021_CVPR}, e.g., batch normalization fusion.
In this case, the neural architecture of LoRA-C CNN model for inference is exactly the same as the pre-trained one.
Therefore, there is not any degradation of IoT quality of services in terms of inference latency.

\section{Experiments} \label{ref-experiments}

This section introduces our experimental setup, including this work's datasets, baselines, and implementation details.
Then, we conduct experimental comparisons and present analyses.
 
\subsection{Experimental Setup}

\begin{table*}[ht]
\centering
\caption{Comparison between training from scratch (SCR), full fine-tuning (FT), and LoRA-C on CIFAR-10/-100 datasets. Our results are highlighted with shading. \#P and \#P (LoRA-C.) refer to the number of parameters that need to be updated when training models and when using our proposed method, respectively. Our method achieves better performance with less number of updated parameters.}
\label{tab:cifar10vscifar100}
\resizebox{0.98\textwidth}{!}{%
\begin{tabular}{lccccccccc}
\toprule
\multirow{3}*{\makecell[l]{Models}} & & \multicolumn{4}{c}{{CIFAR-10}} & \multicolumn{4}{c}{{CIFAR-100}} \\ \cmidrule(lr){3-6} \cmidrule(lr){7-10} 
 &\#P  &\cellcolor{gray!20}\#P (LoRA-C.)  &Acc. (SCR/FT) & \cellcolor{gray!20}Acc. (LoRA-C.) &$\Delta_{Acc}(\uparrow)$   &\cellcolor{gray!20}\#P (LoRA-C.) & Acc. (SCR/FT) & \cellcolor{gray!20}Acc. (LoRA-C.) &$\Delta_{Acc}(\uparrow)$ \\ \midrule
 
ResNet-18 & $11.23$ &\cellcolor{gray!20}$4.26$ & $95.45/94.96$ & \cellcolor{gray!20}$95.69$ & ${\bf{0.24}}$ & \cellcolor{gray!20}$4.26$ & $76.88/76.55$ & \cellcolor{gray!20}$79.93$ & ${\bf{3.05}}$ \\ 

ResNet-34 & 21.29 &\cellcolor{gray!20}0.29 &95.08/95.48 & \cellcolor{gray!20}96.06 & {\bf{0.98}}  & \cellcolor{gray!20}2.16 & 78.38/78.12 &\cellcolor{gray!20} 82.49 & {\bf{4.11}} \\  

ResNet-50 & 23.52 &\cellcolor{gray!20}3.61 & 95.34/95.70 & \cellcolor{gray!20}96.59 & {\bf{1.25}} & \cellcolor{gray!20}2.26 & 77.69/78.68 & \cellcolor{gray!20}82.98 & {\bf{5.29}}\\ 

ResNet-101 & 42.51 &\cellcolor{gray!20}1.95 & 95.77/95.25 & \cellcolor{gray!20}97.10 & ${\bf{1.33}}$  & \cellcolor{gray!20}4.49 & 79.64/78.65 & \cellcolor{gray!20}84.52 & {\bf{4.88}}\\ \bottomrule
\end{tabular}%
}
\end{table*}

\noindent\textbf{Datasets.} 
We conduct experiments on four benchmark datasets, i.e., CIFAR-10, CIFAR-100, CIFAR-10-C and Icons-50 datasets.
CIFAR-10 and CIFAR-100 are two standard datasets, and CIFAR-10-C and Icons-50 are used as corrupted datasets.
Among them, noise is introduced into the validation set of CIFAR-10-C (see Fig.~\ref{fig:CIFAR-10-C-data} for a visualization), and the styles of the training data and test data of Icons-50 are pretty different (see Fig.~\ref{fig:Icon-50-dataset} for a visualization).
\begin{itemize}
    \item \underline{CIFAR-10 and CIFAR-100}~\cite{Alex@Learning} are multi-class natural object datasets for image classification. 
They consist of 50, 000 training images and 10,000 test images in 10/100 categories. Each image has a resolution of $32\times32$ pixels.

\item \underline{CIFAR-10-C}~\cite{Hendrycks@Benchmarking} is a test dataset after using synthetic common perturbations and noise corruptions on the CIFAR-10 test set. It consists of 10, 000 test images of 19 types of damage in 4 categories. The resolution of each image is $32\times32$ pixels.

\item \underline{Icons-50}~\cite{Hendrycks@Benchmarking} consists of 10,000 images in 50 categories collected from different companies. 
We set the resolution of all images to $32\times32$ pixels. 
For training, the data of one company is retained as test data, and the data of the rest of the companies is used as training data.
\end{itemize}

Please refer to~\cite{Ding@Towards} for more explanation of the CIFAR-10-C and Icons-50 datasets.

\noindent\textbf{Baselines.}  
We use RestNet series methods as the baselines, including ResNet-18/-34/-50/-101~\cite{resnet}.

\noindent\textbf{Implementation Details.}  
We implement our method in Python 3.8 and PyTorch 2.2 with CUDA 12.1. All experiments are performed on NVIDIA RTX 4090 GPUs.
We use the pre-trained weights of the ResNet series (i.e., ResNet-18/-34/-50/-101) on ImageNet-1K provided by PyTorch and replace the first $7\times 7$ convolution in ResNet with a $3\times 3$ convolution to process $32\times 32$ image data.
In addition, the output head is modified for different datasets.
When fine-tuning the model, the updating parameters include the replaced first layer $3\times 3$ convolution, the last fully connected layer, the added $\bm{A}$ and $\bm{B}$ matrices, and other parameters are frozen by removing gradients.
We set $r$ to $\{1, 2, 4, 8, 16, 32, 64\}$ and $\alpha$ to $\{1, 2, 4, 8, 16, 32, 64, 128\}$ for traversal.
We use the SGD optimizer for all datasets and set the weight decay to $5e^{-4}$ and the initial learning rate to $0.1$.

\subsection{Experimental Results}

\subsection{The amount of Fine-Tuning Parameters}
\begin{figure}[t]
\centering
\includegraphics[width=1\linewidth]{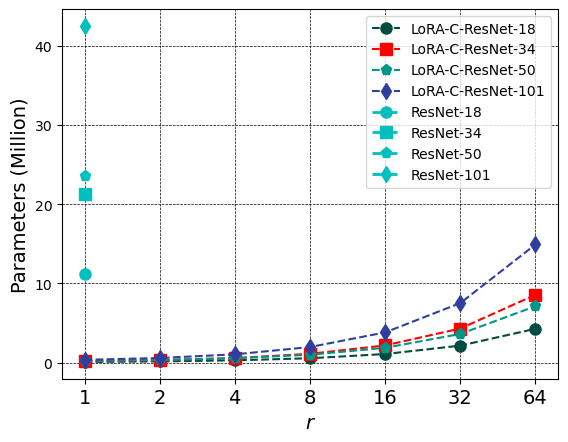}
\caption {Comparison of the updated parameters of \ourmethod{}s and the standard models.}
\label{fig:Updatedvsparameters} 
\end{figure}
The proposed \ourmethod{} reduces the number of model updated parameters by one order of magnitude compared to fine-tuning the full model parameters.
As shown in Fig.~\ref{fig:Updatedvsparameters}, compared with the full parameter fine-tuning ResNet-18, the amount of fine-tuning parameters of \ourmethod{}-ResNet-18 is reduced by up to about 99.35\%.
Compared with the full parameter fine-tuning ResNet-101, the amount of fine-tuning parameters of \ourmethod{}-ResNet-101 is reduced by up to about 99.21\%.
\ourmethod{}-based fine-tuning freezes the backbone model parameters and only fine-tunes the newly added parameters.
The number of newly added parameters is small because \ourmethod{} performs low-rank decomposition in units of convolutional layers, thus effectively reducing the number of model fine-tuning parameters.

\subsection{Results on Standard Benchmarks}

The proposed \ourmethod{} achieves improved model performance with only a few parameters fine-tuned.
As shown in Table~\ref{tab:cifar10vscifar100}, on the CIFAR-10 dataset, based on ResNet-18, our proposed \ourmethod{} achieves 95.69\%  accuracy, surpassing the training from scratch (SCR) result by +0.24\%. 
On the CIFAR-100 dataset, our proposed \ourmethod{} achieves 79.93\% accuracy, surpassing the SCR result by +3.05\%.
This is mainly because \ourmethod{} relies on the existing knowledge learned by the model on ImageNet.
When the main model is frozen, and only the newly added parameters are updated, it is equivalent to allowing the new branches to learn specific features based on the knowledge learned by the model on ImageNet.
Therefore, the \ourmethod{} can achieve higher accuracy than training the model from scratch.

\begin{table*}[t]
    \centering
    \caption{Robustness to common observable corruptions. SCR means the model is trained from scratch. FT means fine-tuning all parameters using new data based on the pre-trained model. Our results are highlighted with shading.}
    \label{table:CIFAR-10-C}
    \subfloat[ResNet18]{
        \scalebox{1.2}{
            \begin{tabular}{lcccccc} \toprule
             Method&  Noise& Blur& Weather&Digital &mean &$\Delta_{Acc}(\uparrow)$ \\ \hline
             SCR & $56.48$& $71.88$& $85.20$& $80.49$&$ 73.51$ &--\\ 
             FT & $55.51$& $70.29$& $84.14$& $80.26$&$ 72.55$ &-$0.96\%$\\ 
             \cellcolor{gray!20}\ourmethod{} (Ours)& \cellcolor{gray!20}$56.07$& \cellcolor{gray!20}$76.89$& \cellcolor{gray!20}$85.47$& \cellcolor{gray!20}$82.11$&\cellcolor{gray!20}\bm{$75.14$} &\cellcolor{gray!20}\textcolor{red}{+$1.63\%$}\\ \bottomrule
            \end{tabular}
        }
        
    }\\ 
    \vspace{5pt}
    \subfloat[ResNet34]{
        \scalebox{1.2}{
            \begin{tabular}{lcccccc} \toprule
             Method&  Noise& Blur& Weather& Digital &mean &$\Delta_{Acc}(\uparrow)$\\ \hline
             SCR& $58.11$& $71.32$& $85.49$& $80.53$&$73.86$ &--\\ 
             FT& $59.1$& $72.33$& $85.82$& $81.18$&$74.61$ &\cellcolor{gray!20}\textcolor{red}{+$0.75\%$}\\ 
             \cellcolor{gray!20}\ourmethod{} (Ours)& \cellcolor{gray!20}$66.08$& \cellcolor{gray!20}$81.06$& \cellcolor{gray!20}$87.69$& \cellcolor{gray!20}$84.22$&\cellcolor{gray!20}\bm{$79.76$} &\cellcolor{gray!20}\textcolor{red}{+$5.9\%$}\\ \bottomrule
            \end{tabular}
        }
    }\\ 
        \vspace{5pt}
    \subfloat[ResNet50]{
        \scalebox{1.2}{
            \begin{tabular}{lcccccc} \toprule
             Method&  Noise& Blur& Weather& Digital &mean &$\Delta_{Acc}(\uparrow)$\\ \hline
             SCR& $53.81$& $70.56$& $85.70$& $81.63$&$72.93$ &--\\ 
             FT& $54.92$& $71.35$& $85.87$& $81.17$&$73.33$ &\cellcolor{gray!20}\textcolor{red}{+$0.4\%$}\\ 
             \cellcolor{gray!20}\ourmethod{} (Ours)& \cellcolor{gray!20}$66.76$& \cellcolor{gray!20}$81.94$& \cellcolor{gray!20}$89.82$& \cellcolor{gray!20}$84.42$&\cellcolor{gray!20}\bm{$80.74$} &\cellcolor{gray!20}\textcolor{red}{+$7.81\%$}\\ \bottomrule
            \end{tabular}
        }
    }\\ 
    \vspace{5pt}
    \subfloat[ResNet101]{
        \scalebox{1.2}{
            \begin{tabular}{lcccccc} \toprule
             Method&  Noise& Blur& Weather& Digital &mean &$\Delta_{Acc}(\uparrow)$\\ \hline
             SCR &$55.10$& $73.03$& $85.30$& $82.35$&$73.95$ &--\\ 
             FT&$56.22$& $73.48$& $85.61$& $82.05$&$74.34$ &\cellcolor{gray!20}\textcolor{red}{+$0.39\%$}\\ 
             \cellcolor{gray!20}\ourmethod{} (Ours)&\cellcolor{gray!20}$73.78$& \cellcolor{gray!20}$82.87$& \cellcolor{gray!20}$90.58$& \cellcolor{gray!20}$86.53$&\cellcolor{gray!20}\bm{$83.44$} &\cellcolor{gray!20}\textcolor{red}{+$9.50\%$}\\ \bottomrule
            \end{tabular}
        }
    }     
\end{table*}

The proposed \ourmethod{} can improve the model performance when the model accuracy is relatively low.
As shown in Table~\ref{tab:cifar10vscifar100}, our proposed \ourmethod{} significantly outperforms SCR. 
On the CIFAR-100 dataset, based on ResNet-34, our proposed \ourmethod{} surpasses the SRC by +4.11\%. Based on ResNet-50, our proposed \ourmethod{} surpasses the SCR by +5.29\%.
The low performance of the model indicates that it is difficult for the model to extract effective features from the current dataset.
\ourmethod{} can achieve high accuracy because it uses a pre-trained model, that is, it extracts effective features based on the existing knowledge of the pre-trained model.
It can also be understood that based on the pre-trained model, the newly added network branch is easier to update parameters through gradient descent.

As the model capacity increases, the model accuracy improves more significantly.
As shown in Table~\ref{tab:cifar10vscifar100}, on the CIFAR-10 dataset, based on ResNet-18, \ourmethod{} improves accuracy by 0.24\% compared to SCR. Based on ResNet-34, \ourmethod{} improves accuracy by 0.98\% compared to SCR. Based on ResNet-101,  \ourmethod{} improves accuracy by 1.33\% compared to SCR.
Compared with simple-structured models processing complex data, complex-structured models can achieve higher performance when processing complex data.
Complex models are more likely to fall into the gradient vanishing problem, which makes it impossible to effectively update model parameters.
However, once the parameters can be effectively updated through gradient descent, their performance often exceeds that of simple models.
\ourmethod{} uses the pre-trained model to update parameters so that the newly added branches can effectively use gradient descent to update their parameters, thereby significantly improving the performance of large-capacity CNN models.

The full fine-tuning (FT) and SCR have similar accuracy. In some cases, the accuracy of FT is even lower than that of SCR.
For example, based on ResNet-18, on CIFAR-10 dataset, The accuracy of SCR is higher than that of FT.
Based on ResNet-34, the accuracy of SCR is lower than that of FT.
This may be due to the overfitting problem of FT.

\subsection{Results on Robustness}

\subsubsection{Limited Training data}
\begin{figure}
\centering
\includegraphics[width=1\linewidth]{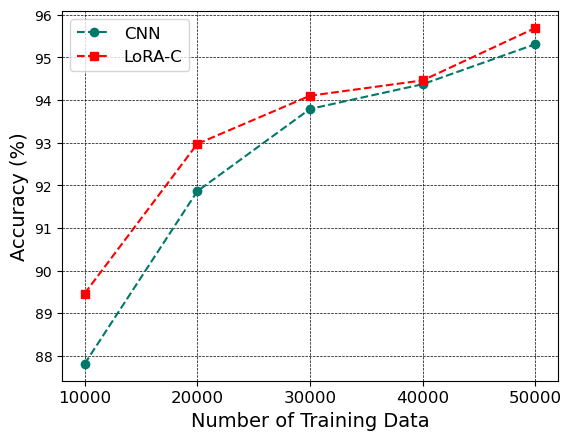}
\caption {Robustness to limited training data.}
\label{fig:limited_data} 
\end{figure}

\ourmethod{} outperforms standard CNN with limited training data, as shown in Fig.~\ref{fig:limited_data}.
The backbone network is learned on the ImageNet dataset.
According to the characteristics of the CNN model, learning on this basis is conducive to the gradient transfer of the model. It is more conducive to obtaining parameters enabling the model to perform well.
\subsubsection{Corrupted Data} \label{results: corrupted data}
\begin{table*}[t]
    \centering
    \caption{Robustness to different styles. Our results are highlighted with shading.}
    \label{table:Icons50}
    \subfloat[ResNet18]{
        \scalebox{1.2}{
            \begin{tabular}{lcccccc} \toprule
             Method&  Apple& Facebook& Google&Samsung&mean &$\Delta_{Acc}(\uparrow)$ \\ \hline
             SCR&$93.99$ &$90.82$ &$85.75$& $85.71$&$89.07$  &-- \\ 
             FT&$95.32$ &$91.11$ &$86.85$& $86.52$&$89.95$  &\cellcolor{gray!20}\textcolor{red}{+$0.88\%$}\\ 
             \cellcolor{gray!20}\ourmethod{} (Ours)&\cellcolor{gray!20}$98.57$&\cellcolor{gray!20}$95.98$&\cellcolor{gray!20} $94.49$&\cellcolor{gray!20} $93.51$&\cellcolor{gray!20}\bm{$95.64$}&\cellcolor{gray!20}\textcolor{red}{+$6.57\%$}\\
             \bottomrule
            \end{tabular}
        }
    }\\ 
    \vspace{5pt}
    \subfloat[ResNet34]{
       \scalebox{1.2}{
            \begin{tabular}{lcccccc} \toprule
             Method&  Apple& Facebook& Google&Samsung&mean &$\Delta_{Acc}(\uparrow)$\\ \hline
             SCR&$93.79$& $88.19$& $86.61$& $85.11$&$88.42$&--\\ 
             FT&$95.70$& $90.44$& $86.38$& $83.79$&$89.08$&\cellcolor{gray!20}\textcolor{red}{+$0.66\%$}\\ 
             \cellcolor{gray!20}\ourmethod{} (Ours)&\cellcolor{gray!20}$99.14$&\cellcolor{gray!20} $97.32$&\cellcolor{gray!20} $96.69$&\cellcolor{gray!20} $94.43$&\cellcolor{gray!20}\bm{$96.90$}&\cellcolor{gray!20}\textcolor{red}{+$8.48\%$}\\
             \bottomrule
            \end{tabular}
        }
    }    
\end{table*}

\begin{figure*}[t]
    \centering
    \begin{subfigure}[b]{0.33\textwidth}
    \centering
    \includegraphics[width=1\textwidth]{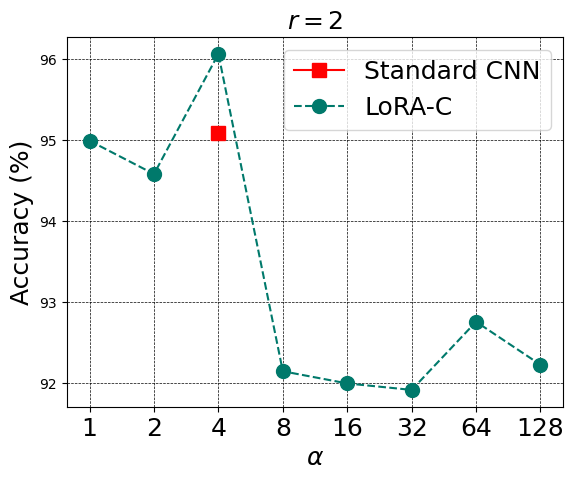}
    \end{subfigure}\hfill
    \centering
    \begin{subfigure}[b]{0.33\textwidth}
    \centering
    \includegraphics[width=1\textwidth]{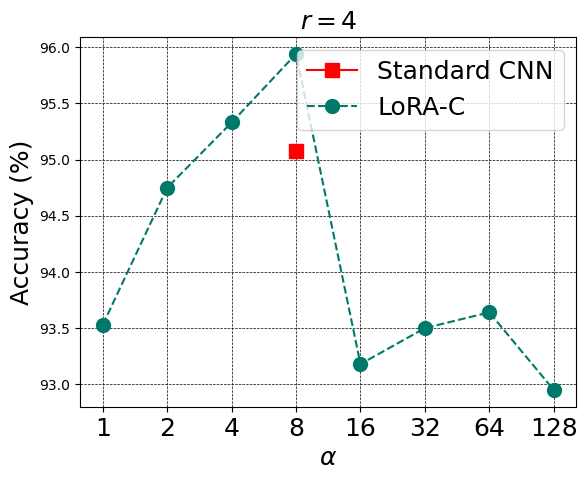}
    \end{subfigure}\hfill
    \centering
    \begin{subfigure}[b]{0.33\textwidth}
    \centering
    \includegraphics[width=1\textwidth]{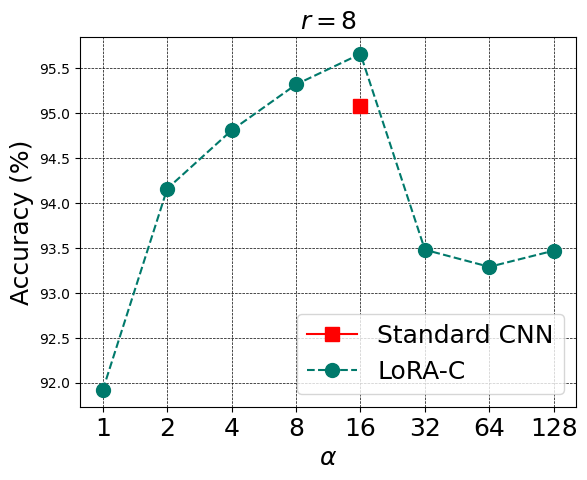}
    \end{subfigure}\hfill
    \centering
    \begin{subfigure}[b]{0.33\textwidth}
    \centering
    \includegraphics[width=1\textwidth]{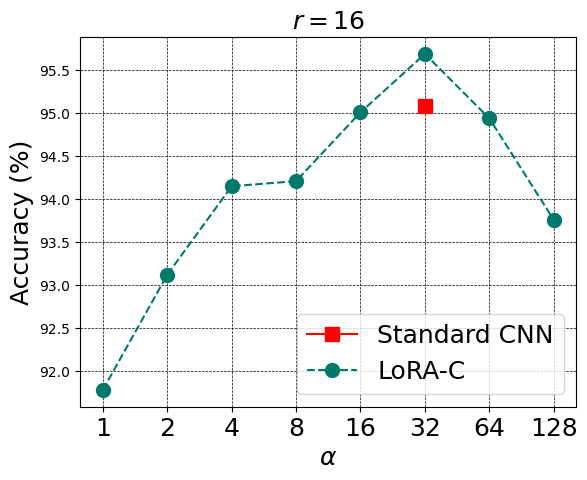}
    \end{subfigure}
    \centering
    \begin{subfigure}[b]{0.33\textwidth}
    \centering
    \includegraphics[width=1\textwidth]{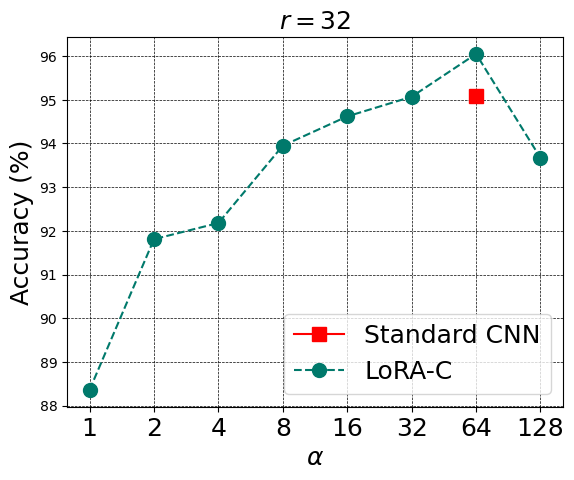}
    \end{subfigure}\hfill
    \centering
    \begin{subfigure}[b]{0.33\textwidth}
    \centering
    \includegraphics[width=1\textwidth]{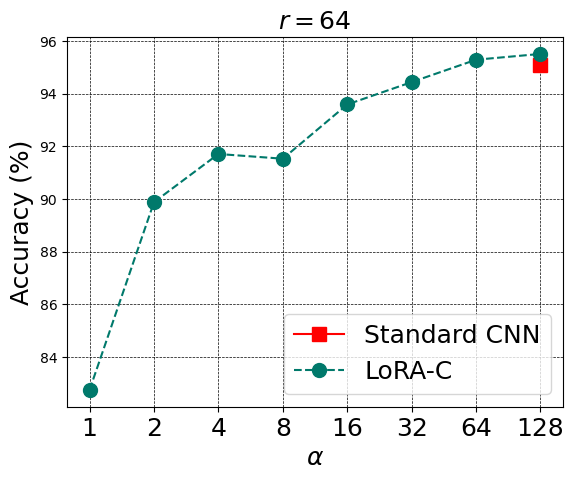}
    \end{subfigure} 
    \caption{Given $r$, the impact of $\alpha$ on the CIFAR-10 dataset.\label{fig:cifar10-resnet34}}
\end{figure*}
\begin{figure*}[t]
    \centering
    \begin{subfigure}[b]{0.33\textwidth}
    \centering
    \includegraphics[width=1\textwidth]{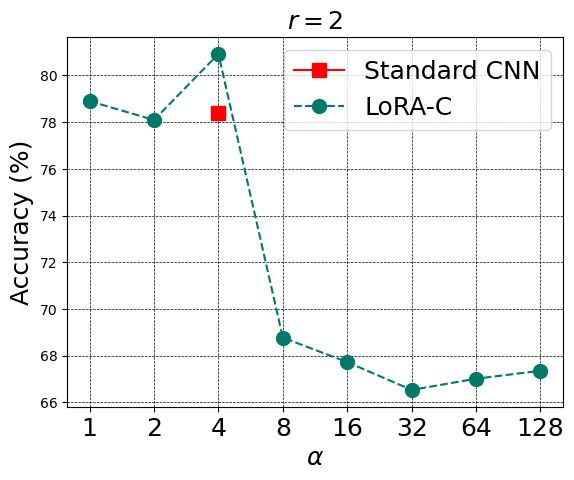}
    \end{subfigure}\hfill
    \centering
    \begin{subfigure}[b]{0.33\textwidth}
    \centering
    \includegraphics[width=1\textwidth]{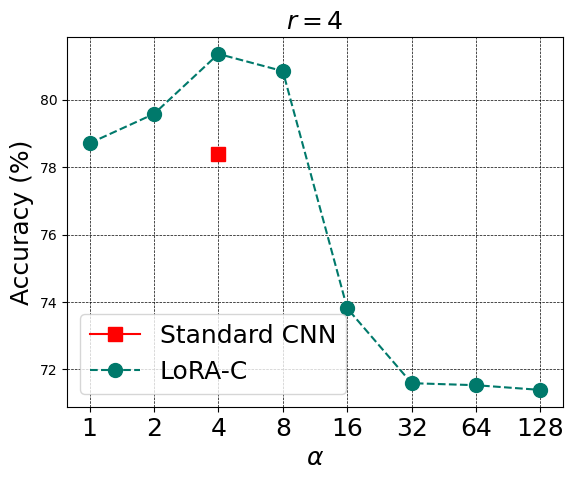}
    \end{subfigure}\hfill
    \centering
    \begin{subfigure}[b]{0.33\textwidth}
    \centering
    \includegraphics[width=1\textwidth]{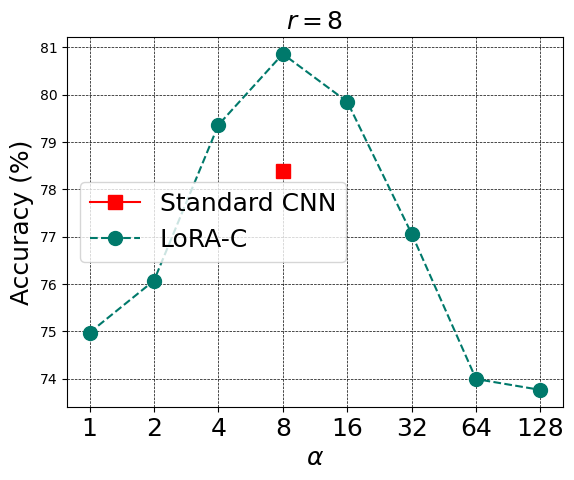}
    \end{subfigure}\hfill
    \centering
    \begin{subfigure}[b]{0.33\textwidth}
    \centering
    \includegraphics[width=1\textwidth]{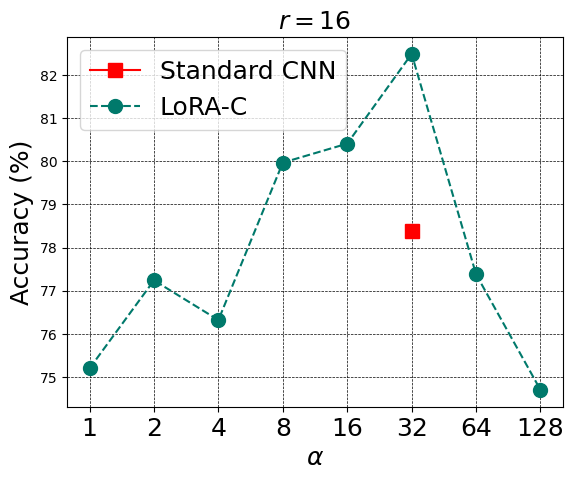}
    \end{subfigure}
    \centering
    \begin{subfigure}[b]{0.33\textwidth}
    \centering
    \includegraphics[width=1\textwidth]{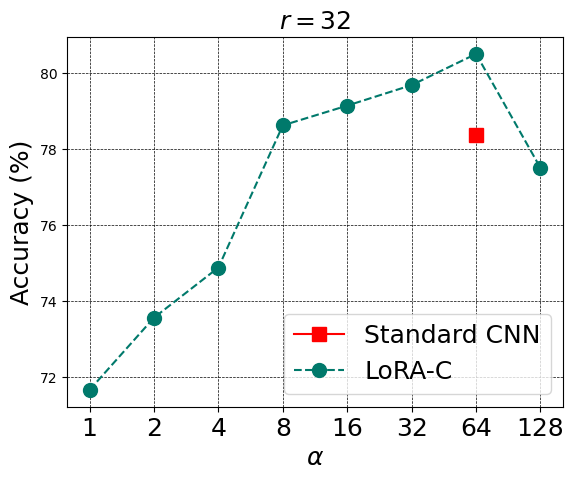}
    \end{subfigure}\hfill
    \centering
    \begin{subfigure}[b]{0.33\textwidth}
    \centering
    \includegraphics[width=1\textwidth]{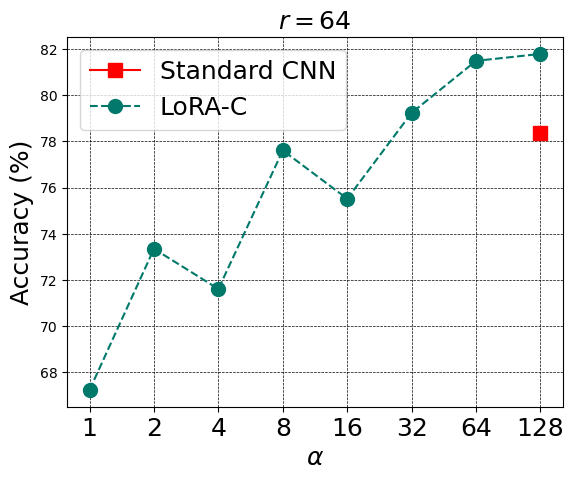}
    \end{subfigure} 
    \caption{Given $r$, the impact of $\alpha$ on the CIFAR-100 dataset.\label{fig:cifar100-resnet34}}
\end{figure*}

The proposed \ourmethod{} significantly improves the performance of handling corrupted data.
As shown in Table~\ref{table:CIFAR-10-C}, based on ResNet-34, compared to the accuracy of SCR, \ourmethod{} has an accuracy improvement of 5.9\%.
Based on ResNet-50, compared to the accuracy of SCR, \ourmethod{} has an accuracy improvement of 7.81\%.
It is worth mentioning that the \ourmethod{}-based fine-tuning method achieves better performance in all categories, as shown in Table~\ref{table:CIFAR-10-C}.
In \ourmethod{}, the backbone network parameters are frozen and will not be updated with local data.
The parameters of the backbone network are obtained based on ImageNet training, which means that the backbone network retains the knowledge learned on ImageNet.
On the one hand, based on the backbone network, the knowledge learned by the backbone network on ImageNet can improve performance by using local data to fine-tune the newly added model parameters.
On the other hand, the parameters of the backbone model are frozen, which can play a regularization role.
Therefore, the proposed \ourmethod{} can achieve better results on the corrupted dataset that is, \ourmethod{} has strong robustness.

\subsubsection{Data Under Different Styles}
The proposed \ourmethod{} achieves high performance when dealing with different training and test data styles.
As shown in Table~\ref{table:Icons50}, based on ResNet-18, compared to the accuracy of SCR, \ourmethod{} has an accuracy improvement of 6.57\%.
Based on ResNet-34, compared to the accuracy of SCR, \ourmethod{} has an accuracy improvement 8.48\%.
In addition, similar to those obtained with corrupted data, the \ourmethod{}-based fine-tuning method achieves better performance in all categories.
The reasons for this are the same as in Section~\ref{results: corrupted data}:
(i) Learning is performed based on the existing knowledge of the backbone network.
(ii) The parameters of the backbone model are frozen, regularizing the model with the newly added parameters.
The above results prove that the proposed \ourmethod{} is highly robust.

\subsection{Hyperparameter Study} \label{ref-hyperparameter}

\begin{figure}
\centering
\includegraphics[width=0.9\linewidth]{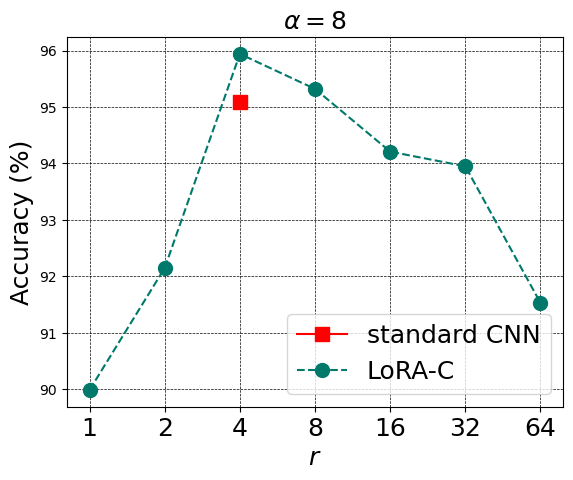}
\caption {Given $\alpha=8$, the impact of $r$ on CIFAR-10 dataset.}
\label{fig:a_8_cifar10}
\end{figure}
\begin{figure}
\centering
\includegraphics[width=0.9\linewidth]{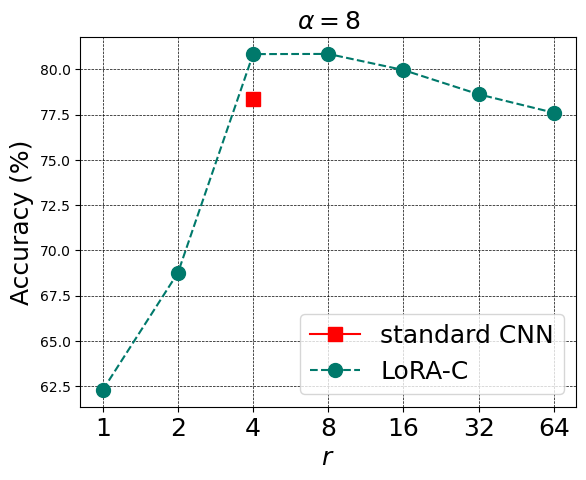}
\caption {Given $\alpha=8$, the impact of $r$ on CIFAR-100 dataset.}
\label{fig:a_8_cifar100}
\end{figure}

\begin{figure*}[t]
    \centering
    \begin{subfigure}[b]{0.33\textwidth}
    \centering
    \includegraphics[width=1\textwidth]{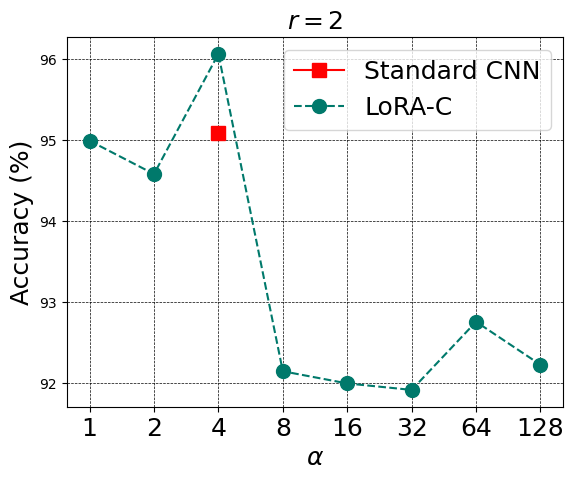}
    \end{subfigure}\hfill
    \centering
    \begin{subfigure}[b]{0.33\textwidth}
    \centering
    \includegraphics[width=1\textwidth]{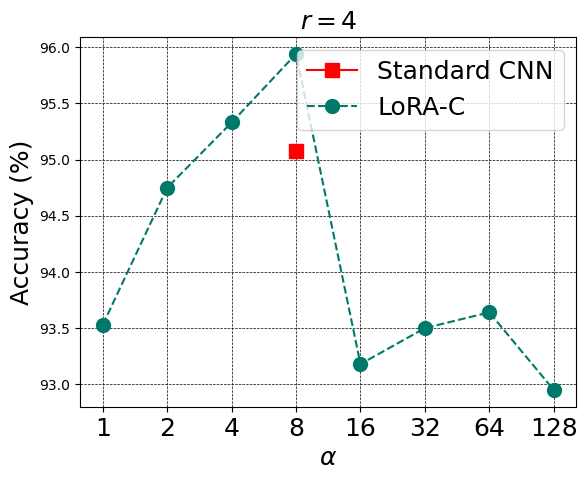}
    \end{subfigure}\hfill
    \centering
    \begin{subfigure}[b]{0.33\textwidth}
    \centering
    \includegraphics[width=1\textwidth]{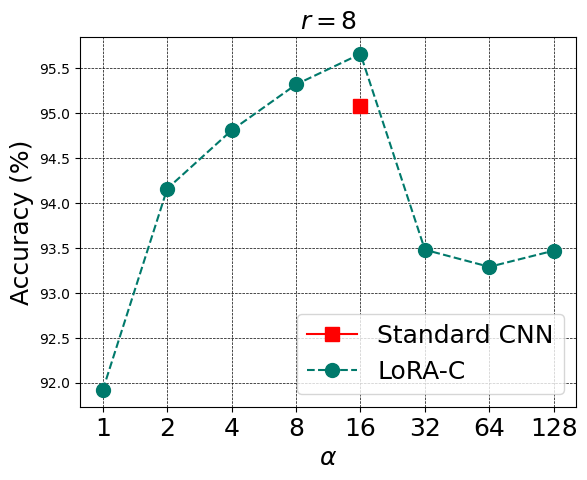}
    \end{subfigure}
    \caption{Given $\alpha$, the impact of $r$ on the CIFAR-10-C dataset.\label{fig:cifar10-C-resnet18}}
\end{figure*}

Two hyperparameters have a great impact on the \ourmethod{}, namely $\alpha$ and $r$.
The $\alpha$ measures the proportion of newly added branches compared to the backbone network.
The larger the $\alpha$, the greater the proportion of newly added branches, and vice versa.
The $r$ represents the rank of $\Delta W$. The larger $r$ is, the more parameters are fine-tuned, and vice versa.

Figs.~\ref{fig:cifar10-resnet34} and~\ref{fig:cifar100-resnet34} illustrate the relationship between model accuracy and $\alpha$ for a given $r$.
We observe that the performance of the \ourmethod{} does not always improve with the increase of $\alpha$.
For example, when $r=2$, \ourmethod{} achieves the highest accuracy when $\alpha=4$.
When $r=4$, \ourmethod{} achieves the highest accuracy when $\alpha=8$.
When $r=64$, \ourmethod{} achieves the highest accuracy when $\alpha=128$.
We also observe that the best performance is usually obtained when $\frac{\alpha}{r} = 2$, as shown in Figs.~\ref{fig:cifar10-resnet34} and~\ref{fig:cifar100-resnet34}.

We also illustrate the relationship between model accuracy and $r$ for a given $\alpha$, as shown in Fig.~\ref{fig:a_8_cifar10} and Fig.~\ref{fig:a_8_cifar100}.
We observe that (i) the best performance is usually obtained when $\ \frac {\alpha}{r} = 2$.
(ii) When fine-tuning, it is not the case that the more parameters updated, the better.

In addition, We also illustrate the relationship between model accuracy and $r$ for a given $\alpha$ on CIFAR-10-C dataset, as shown in Fig.~\ref{fig:cifar10-C-resnet18}.
The best performance is usually obtained when $\ \frac {\alpha}{r} = 2$.
This discovery provides experience for the widespread application of \ourmethod{}.

\section{Conclusion and Future Work} \label{ref-conclusion}
This paper proposes a fine-tuning method for robust CNNs for IoT devices, \ourmethod{}, which performs low-rank decomposition in convolutional layers to reduce the number of fine-tuning parameters.
By setting the ratio of $\alpha$, which controls the proportion of newly added branches, to the rank of the weight matrix to a constant, the accuracy of the fine-tuned model significantly exceeds that of the fully trained model.
Experimental results on CIAFR-10, CIFAR-100, CIFAR-10-C, and Icons50 datasets demonstrate the effectiveness of the proposed \ourmethod{}.  

Given that \ourmethod{} effectively improves the robustness of the CNN model, this motivates our future work to start from the following two points:
(i) Set $\alpha$ to be learnable, and it will learn the optimal value according to the value of $r$ during model training.
(ii) Apply LoRA to MonoCNN or SineFM to explore how to use LoRA to fine-tune the CNN model based on nonlinear mapping.

\bibliographystyle{IEEEtran}
\bibliography{main}
\end{document}